# Scanning Raman spectroscopy for characterizing compositionally spread films

*Venimadhav. A\*, Yates K. A., Blamire M. G.*

Department of Materials Science, University of Cambridge, Pembroke Street, Cambridge, CB2 3QZ, UK

**Abstract**

Composition-spread $La_{1-x}Sr_xMnO_3$ thin films were prepared by pulsed laser deposition technique from $LaMnO_3$ and $SrMnO_3$ targets. The films were epitaxial with a continuous variation of the out of plane lattice parameter along the direction of composition gradient. Scanning Raman spectroscopy has been employed as a non-destructive tool to characterize the composition-spread films. Raman spectra showed the variation of the structural, Jahn Teller distortions and the presence of coexisting phases at particular compositions that are in agreement with the previous observation on the single crystal samples. Raman spectra on the continuous composition-spread film also reveal the effect of disorder and strain on the compositions.

\* Corresponding author: A. Venimadhav: avm24@cus.cam.ac.uk

**Introduction**

Perovskite oxides with strong electron correlation exhibit very complex phase diagrams. The doped lanthanum manganites in particular display a variety of phases including colossal magnetoresistance, charge-ordering at definite doping levels and evidence for the coexistence of different phases at a given composition[1]. The Zener double exchange mechanism and the electron lattice coupling due to the Jahn Teller (JT) effect together control the transport and optical properties of these materials [2]. There is an increasing technological interest in the optimally doped regions of the phase diagram where these materials exhibit 100% spin polarization due to the development of a half-metallic ground state [3]. Finding the critical doping and phase boundaries from conventional preparation methods is laborious and rather time consuming. For getting an overview of the entire composition of the inorganic materials Briceño et al, have employed the combinatorial process [4]. Continuous variation of manganite compositions in thin film form has been created by combinatorial pulsed laser deposition systems [5,6].

It is difficult to characterize compositionally spread films fully by non-destructive techniques, particularly evaluating stoichiometry and the analysis of impurity phases. Raman spectroscopy can be an ideal tool to find any impurities at micron levels and may be more efficient than x-ray diffraction and energy dispersive analysis techniques [7]. Additionally it provides information about the electronic and lattice processes responsible for the physical behavior of the material. Raman spectroscopy has proven to be a very useful technique for evaluating the oxygen stoichiometry [8] and cation disorder [9] in $YBa_2Cu_3O_{7-\delta}$, thereby helping in understanding of the relationship between the superconducting $T_c$ and the crystal structure. It has also provided essential information about the lattice and JT distortion and orbital ordering in the manganites that occurs at variable doping levels [10-20]. In this paper we demonstrate that scanning Raman spectroscopic technique is an efficient tool for characterizing the continuous composition spread (CCS) film.

**Experiment**

CCS films were deposited by the pulsed laser deposition (PLD) technique. Briceño et al, have employed post annealing conditions for the films deposited at the ambient temperature [4]. Christen et al, have obtained epitaxial continuous phases of perovskite oxides at the optimal growth conditions, there by avoiding extensive post-annealing steps [21]. In this study continuous-spread (LaSr)MnO$_3$ films have been prepared by sequential deposition of sub-monolayers from LaMnO$_3$ (LMO) and SrMnO$_3$ (SMO) targets using a multi target carousel. The laser plume and the substrate were stationary during the complete deposition process. Composition variation is usually achieved by creating thickness gradient of the constituent target materials on the substrate by manipulating the laser plume using shutters or moving masks. In our set up we used a shadow mask plate with specific openings that mask the laser plume for different targets as shown in the Figure 1. This plate is an extension to the target carousel placed at a distance 1 cm above the substrate. Alternately exposing half of the substrate to the plume for each target produces composition gradient. This can easily be extended to more than two targets to create a 2-D composition spread. The complete process has been automated. The spread films were deposited on 10 mm wide LaAlO$_3$ (1 0 0) (LAO) substrate at 790°C in a flowing oxygen gas at a chamber pressure of 100 mTorr. After the deposition the films were then annealed at the deposition temperature for one hour in 1atm O$_2$ pressure.

**Results and Discussion**

Crystallinity and the orientation of the films were characterized using a high resolution four circle X-ray diffractometer. The overall θ-2θ scan showed (0 0 l) orientation of the entire composition spread and found no impurities within the experimental error (5%). Figure 2 shows the contour map of the θ-2θ scans at varying positions on the film that were carried out using a line source focused to 5x0.5 mm using cross slits. The (0 0 2) diffraction peak of the composition spread film moves to a higher 2θ position with increasing the Sr content; this is consistent with the larger c-lattice parameter of LMO. The continuous variation of the out of parameter indicates the change in composition of the spread film. The inset of figure 2 shows the d-values at different positions including the end members LMO and SMO and the nominal composition (x) of the spread film is shown in the upper panel. The observed out of plane values are than their bulk c-parameter; this can be understood from the compressive stress on the film due to the smaller lattice parameter of the LAO substrate (3.789 Å).

The Raman study of the film was carried out in backscattering geometry using a Renishaw Ramanscope 1000 equipped with a CCD detector. An Ar ion laser with a 514.5 nm wavelength was focused at a 2μm spot size on the sample. Raman spectra from the composition spread LSMO sample were collected at different positions by scanning the laser spot along the sample surface. The Raman spectra were divided by Bose-Einstein thermal factor [1- exp (-ℏω/$k_B$T)], where ℏ is the Planks constant, ω is the mode frequency, $k_B$ is the Boltzman constant and T is the temperature at which the Raman spectra were recorded (300 K) [19]. Spectra were recorded at variable laser power to ensure that there were no effects due to sample heating. Figure 3 shows the typical Raman spectra obtained at different compositions along the direction of the composition gradient (from La to Sr rich end). We observed peaks at 436 $cm^{-1}$ and 650 $cm^{-1}$; the peak above 650 $cm^{-1}$ has been resolved into two peaks at 640 $cm^{-1}$ and 680 $cm^{-1}$. At the La rich end Raman spectra were dominated by the 640 $cm^{-1}$ and 680 $cm^{-1}$ peaks; increasing Sr content initiates 436 $cm^{-1}$ peak as shown in the figure 3. A peak at 525 $cm^{-1}$ was not identified.

The Raman spectrum of orthorhombic LMO is dominated by two intense peaks at 500 $cm^{-1}$ and 610 $cm^{-1}$ associated with the JT distortion [10]. For Sr doping, Raman spectra of LSMO with orthorhombic structure (x < 0.17) contain modes due to orthorhombic distortion similar to parent orthorhombic LMO. As the doping increases (x > 0.17) the symmetry of the structure changes to rhombohedral characterized by the appearance of a mode at 436 $cm^{-1}$ [15,16]. Recently however it has been shown that for oxygen rich $LaMnO_{3+z}$ the case is more complicated [11]. Instead of sharp peak at 500 $cm^{-1}$ and 610 $cm^{-1}$, the peaks are broader and shifted to 520 and 640 $cm^{-1}$. However their attribution to a JT distortion mode for 520 $cm^{-1}$ and 640 $cm^{-1}$ feature is kept. Positions of 610 $cm^{-1}$ peak increases in frequency as increasing z in $LaMnO_{3+z}$ [12]. On the basis of theses observations the peaks observed on our films at 640 cm$^{-1}$ and 436 cm$^{-1}$ were attributed due to JT effect and rhombohedral symmetry respectively. The peak at 680 cm$^{-1}$ will be discussed later.

The appearance and intensity variation of the peaks at 640 cm$^{-1}$ and 436 cm$^{-1}$ moving along the direction of composition gradient is shown in the figure 4(a) and (b) respectively. The mode around 640 cm$^{-1}$ decreases in intensity continuously as increasing the Sr content; this is expected as the JT distortion decreases as increasing Sr doping. Doping of Sr into LMO brings in ferromagnetic metallic state from antiferromagnetic insulating state. This transition is accompanied by a structural change from orthorhombic to the rhombohedal with increasing Sr content. The Raman shifts are sensitive to the structural changes; decrease in intensity of the 640 cm$^{-1}$ mode is coincident with the appearance of the mode 436 cm$^{-1}$ associated with the rhombohedral structure. It is interesting to note that for composition x =0.15 on the sample both 640 cm$^{-1}$ and 436 cm$^{-1}$ modes are present, indicating the co existence of the co-operative JT distortion with the rhombohedral structure. This is close to the the phase coexistence predicted for LSMO single crystals at x = 0.17 [23]. Above x = 0.2, the 640 cm$^{-1}$ mode completely disappears. The intensity of the 436 cm$^{-1}$ peak in figure 4(b) shows maximum at x = 0.35. It is worth mentioning that at x=0.33 optimal doping and 100% spin polarization was observed in bulk crystals [3]. This peak disappears for x > 0.6 possibly indicating another structural change. No significant Raman peaks were observed for compositions above 0.6 on the sample. Previously tetragonal structure was reported for Sr above 0.54 in LSMO [22]. However the tolerance factor reaching close to 1.0 may lead to the cubic structure in the thin films that would be Raman inactive and may explain the absence of Raman modes above 0.6. It is clear from the above observation that the intensity variation of the JT distortion, the appearance of the co existing phases and then the rhombohedral phase on the spread film are consistent with the structural and transport change observed in LSMO single crystals [23].

Now we discuss the mode at 680 cm$^{-1}$. This mode has neither been predicted nor observed in orthorhombic LMO [10]. However a lattice dynamical calculation on rhombohedral LMO shows a silent mode at 716 cm$^{-1}$ [11]. A mode at 690 cm$^{-1}$ has previously been observed in the LSMO thin films and single crystals [13-15]. The mode was variously been assigned to an impurity mode [12], partial revealing of the phonon density of states (PDOS) [17] or due to second order Raman scattering process [12]. A systematic variation of a peak at 710 cm$^{-1}$ has been observed for Fe substitution of Mn site in La$_{0.7}$Sr$_{0.3}$MnO$_3$ [20]. Softening of this mode was observed as a consequence of increase in the cell volume (or increase in Mn-O distance) due to Fe substitution. Recently a mode at similar Raman shift was observed in LMO single crystals for hydro static pressure above 7 kPa; hardening of this mode is observed as increasing the pressure due to contracting Mn-O bonds [18]. The present Raman study on the continuous variation of (La/Sr)MnO$_3$ can give more insight on this mode. Figure 4(c) shows variation of the intensity of this mode as increasing Sr. We observed hardening of this mode as the Sr content increases from 680 cm$^{-1}$ to 690 cm$^{-1}$ (for the compositions on the sample from x = 0 to 0.4 as in figure 4 (d)). The intensity variation and the hardening of the mode rules out assigning this mode to an impurity. In the present case deposition in the oxygen condition on a smaller lattice substrate together can influence the Raman spectra. The deposition of the films in oxygen conditions can cause disorder from formation of orthorhombic LMO. The disorder in LMO together with the compressive substrate strain can therefore either partially reveal the PDOS or activating the silent mode. The decrease of the lattice parameter from x-ray diffraction relates to the decrease in the Mn-O distance and is consistent with the hardening of this mode.

In conclusion we prepared composition gradient LSMO films by automated-PLD at optimized deposition conditions. The reported method for obtaining the composite gradient is simple in a manner that it does not involve complicated moving masks or prolonged annealing steps. Raman characterization showed modes due to structural changes and the coexistence of phases at the compositions expected for LSMO. The Raman results therefore complement the x-ray diffraction that showed a continuous change of the out of plane parameter with increasing Sr content. Raman spectra also give useful information about the effects due to disorder and strain. We anticipate that the Raman spectroscopy can be a rapid and non-destructive characterization technique for analyzing compositionally graded films to get an overview on the phase diagrams of complex systems that have varying phase boundaries and coexisting phases.

## Aknowledgement

This work was funded by EPRSC

Figure Captions

Figure 1. Combinatorial PLD set up

Figure 2. Contour map showing the θ-2θ scans at varying positions on the sample. Inset shows the variation of d value with the position with the intended composition on the upper panel.

Figure 3. Typical Raman spectra obtained at different compositions on the sample; Sr composition increases from top to bottom in the panel. '*' represents the first order Raman peak due to LAO substrate

Figure 4. Intensity versus composition plots for modes (a) 640 cm$^{-1}$ (b) 436 cm$^{-1}$ (c) 680 cm$^{-1}$ (d) shows the position of the 680 cm$^{-1}$ mode versus composition.

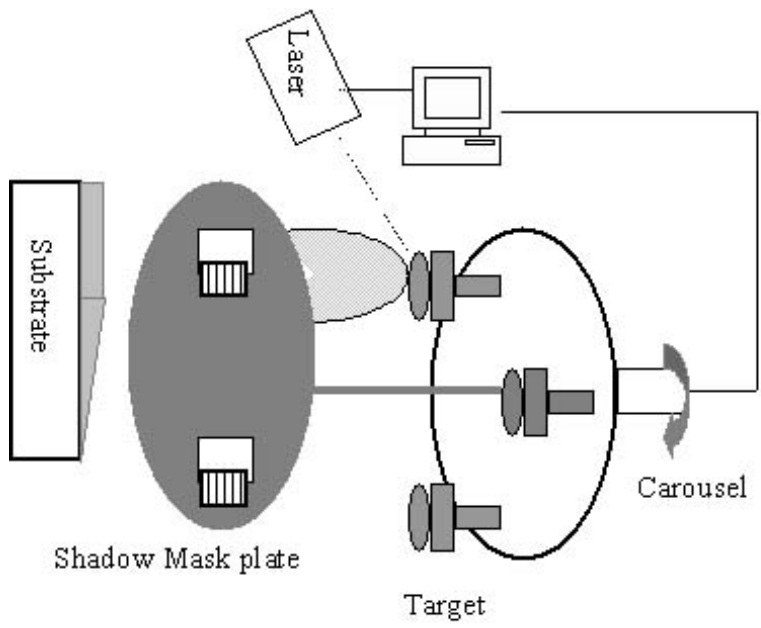

Venimadhav *et al*, Figure 1

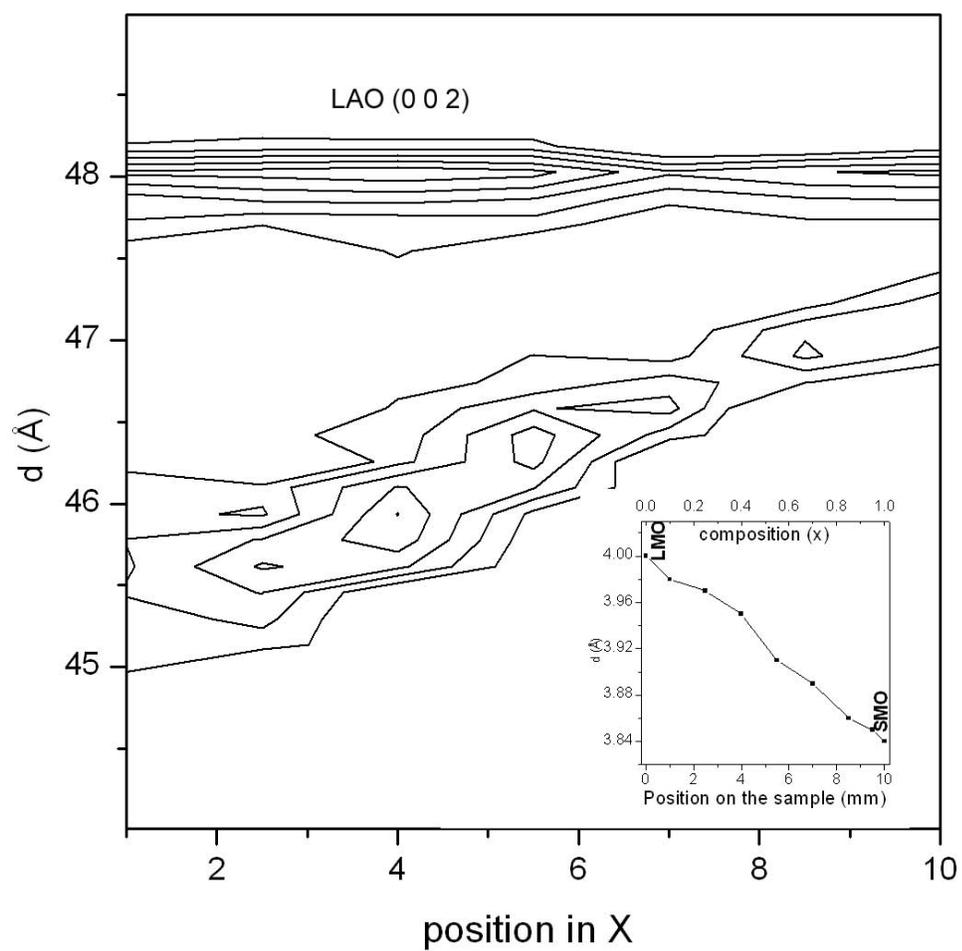

Venimadhav *et al,* Figure 2

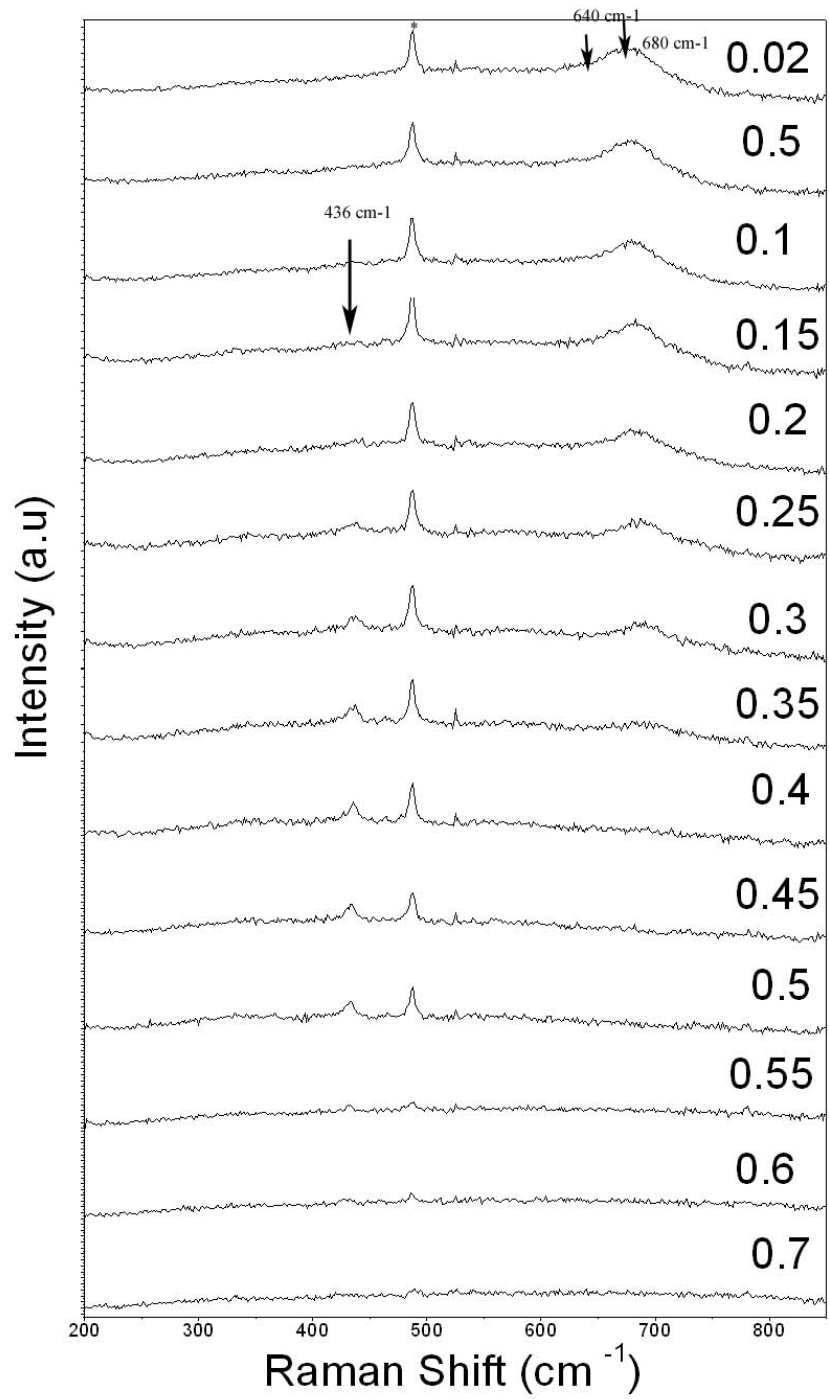

Venimadhav *et al*, Figure 3

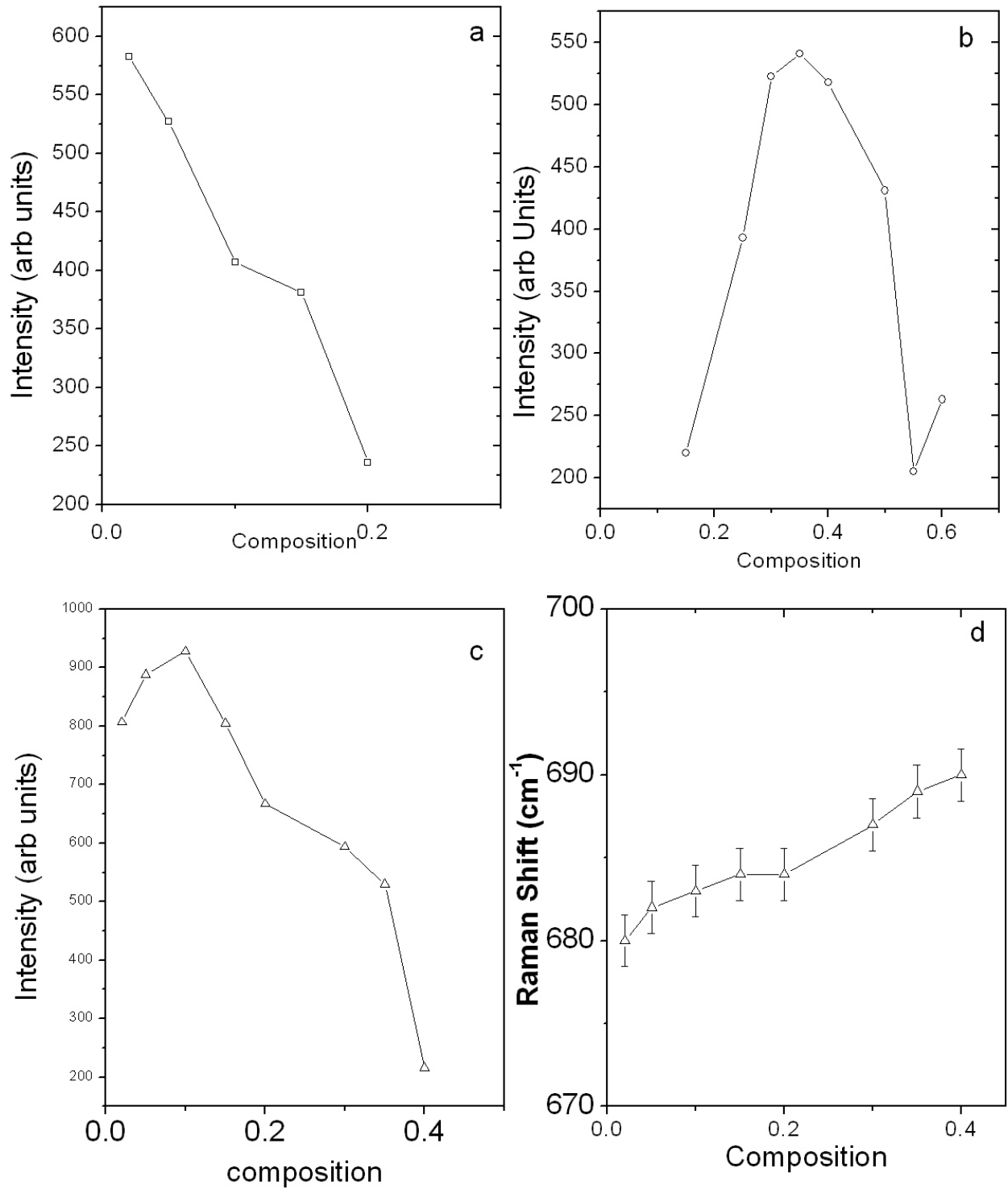

Venimadhav *et al*, Figure 4